\documentclass[aps,prx,twocolumn,amsmath,amssymb,showpacs,superscriptaddress,notitlepage]{revtex4-1}
\usepackage{amsmath,amssymb,amsfonts,bm}
\usepackage{graphicx}
\usepackage{epstopdf}
\usepackage{dcolumn}
\usepackage{mathrsfs}
\usepackage{bbold}
\usepackage{dsfont}
\usepackage{dcolumn}
\usepackage[colorlinks=true,linkcolor=blue,citecolor=blue, urlcolor=blue,bookmarks=false]{hyperref}
\usepackage{changes}
\colorlet{Changes@Color}{red}

\begin{document}

\title{Anomalous Hall effects in magnetic weak topological insulator films}

\author{Rui Chen}
\affiliation{Department of Physics, Hubei University, Wuhan 430062, China}
\author{Xiao-Xia Yi}
\affiliation{Department of Physics, Hubei University, Wuhan 430062, China}
\author{Bin Zhou}
\affiliation{Department of Physics, Hubei University, Wuhan 430062, China}
\affiliation{Key Laboratory of Intelligent Sensing System and Security of Ministry of Education,\\
Hubei University, Wuhan 430062, China}
\author{Dong-Hui Xu}\email[]{donghuixu@cqu.edu.cn}
\affiliation{Department of Physics and Chongqing Key Laboratory for Strongly Coupled Physics, Chongqing University, Chongqing 400044, China}
\affiliation{Center of Quantum Materials and Devices, Chongqing University, Chongqing 400044, China}

\begin{abstract}
The interplay between magnetism and strong topological insulator gives rise to distinct new topological phases and various intriguing phenomena, attracting significant attention in recent years. However, magnetic effects in weak topological insulators remain largely unexplored. In this work, we systematically investigate the magnetic effect on thin films of weak topological insulators. We focus on ferromagnetic and antiferromagnetic effects, which have been extensively studied in strong topological insulators, as well as the recently highlighted altermagnetic effect. We reveal that the interplay between magnetism and weak topological insulators leads to a variety of Hall effects in the absence of an external magnetic field, including the metallic quantum anomalous Hall effect without chiral edge states, the quantum anomalous Hall effect with a higher Hall conductance plateau, the quantized layer Hall effect, the metallic half-quantized valley-like Hall effect, and a quantized valley-like Hall effect.  This work provides valuable insights for exploring magnetic effect on weak topological insulators.
\end{abstract}
\maketitle
\section{Introduction}
Over the past two decades, topological insulators have attracted considerable attention due to their fundamental novelty and potential applications~\cite{Hasan2010RMP,Qi2011RMP,Shen2017TI,Bernevig2013TI}. Unlike conventional insulators, topological insulators exhibit hallmark gapless surface Dirac states with a linear dispersion, resulting from the symmetry-protected nontrivial bulk topology. Within the topological band theory~\cite{FL07PRB,Moore07prbrc,FL07PRL}, three-dimensional topological insulators are
classified into strong topological insulators (STIs) and weak topological insulators (WTIs), distinguished by their odd or even number of surface Dirac cones, respectively. STIs have been experimentally realized in materials like Bi$_{1-x}$Sb$_x$~\cite{Hsieh2009Science}, Bi$_2$Se$_3$~\cite{Zhang09np,Xia09natphys,Chen2009Science,Zhang10njp} family, and TiBiSe$_2$ family~\cite{ChenYL10PRL,Kuroda10PRL,Sato10PRL,Lin10PRL}, while WTIs have emerged in ZrTe$_5$~\cite{Zhang2021Nc}, Bi$_4$I$_4$~\cite{Noguchi2019Nature,Huang2021PRX}, and Bi$_4$Br$_2$I$_2$~\cite{Zhong2023Nc}.

The interplay between magnetism and STIs leads to the discovery of novel topological phases, such as the quantum anomalous Hall insulator~\cite{Chang2013Science,WangHW2023APS}, axion insulator~\cite{Liu20nm,Mogi17nm}, semi-magnetic topological insulator~\cite{mogi2021experimental,BiSH2024arXiv}, metallic quantum anomalous Hall insulator~\cite{Bai2023arXiv}, and half-quantum mirror Hall insulator~\cite{FuB2024NC}.
These phases have been experimentally observed in ferromagnetic topological
insulators like (Bi, Sb)$_2$Te$_3$ and the intrinsic antiferromagnetic topological insulator MnBi$_2$Te$_4$~\cite{Bernevig2022Nat,Tokura19nrp,Wang2021TI,Liu2021AdvMat,XuHK2022QF,TanWL2022QF}. However, the magnetic effect on WTIs remains an unexplored territory. Concurrently, the emergence of altermagnetism, a unique form of collinear antiferromagnetism, has attracted significant attention. This magnetic state possesses the ability to induce a spontaneous Hall effect~\cite{Mejkal2020SA,Libor22PRX,Smejkal22PRX,Feng2022Natelec,Betancourt23PRL}. Moreover, the band topology induced by altermagnetism has been found in different works~\cite{Fernandes2024prb,RaoP2024prb,li2024topological,Zhu2023prb}.

\begin{figure}[tpb]
\centering
\includegraphics[width=\columnwidth]{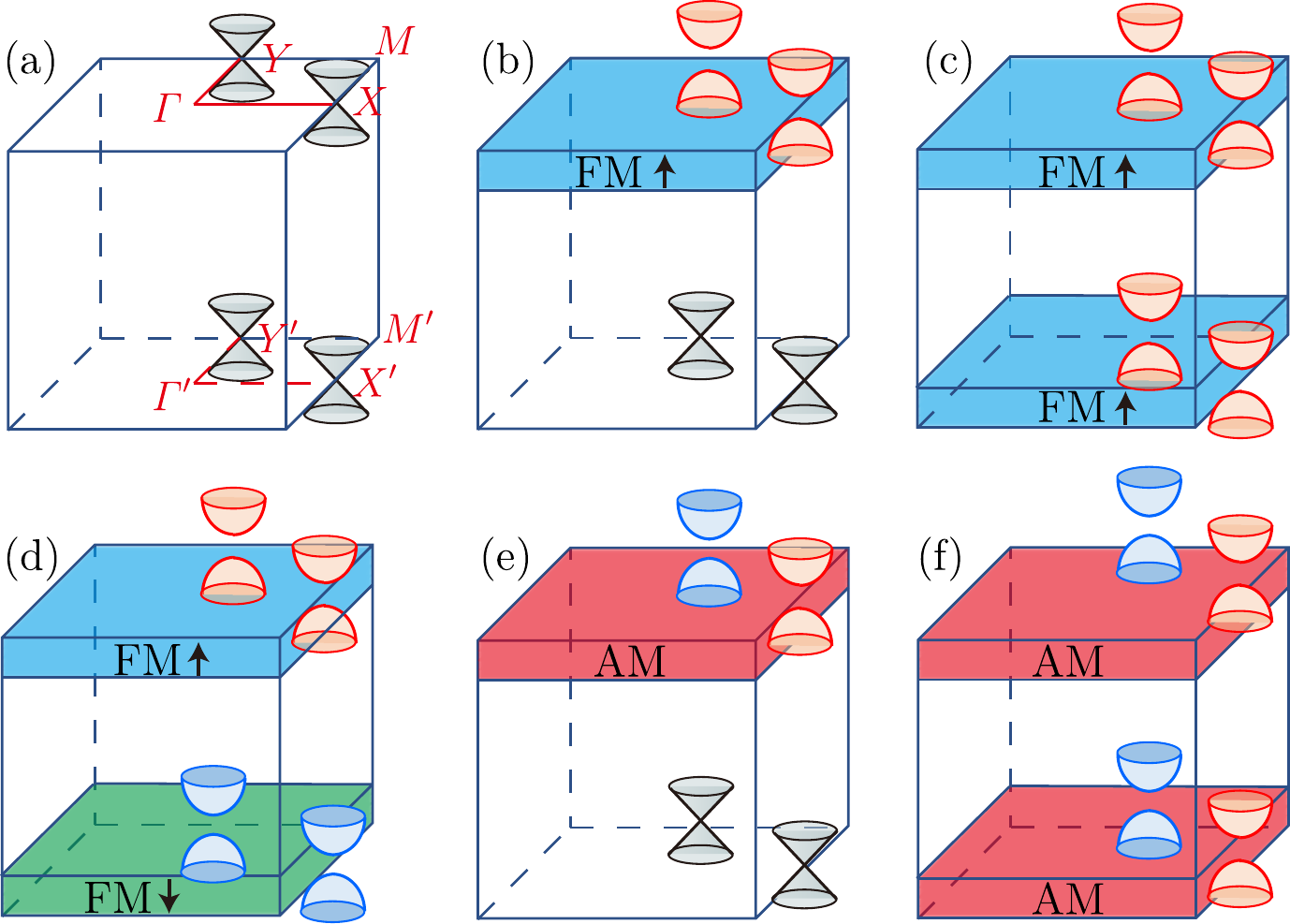}
\caption{Schematic illustrations of six different configurations in the WTI film. (a) The non-magnetic WTI film hosts four gapless Dirac cones. Two of them are located at the $X$ and $Y$ points on the top surface, and another two of them are located at the $X^\prime$ and $Y^\prime$ points on the bottom surface. [(b)-(f)] The surface Dirac cone becomes gapped for the magnetic surface layers. Specifically, the system in (b) realizes the metallic quantized anomalous Hall effect, the system in (c) realizes a quantum anomalous Hall effect with a higher Hall conductance plateau, the system in (d) realizes a quantized layer Hall effect, the system in (e) realizes a metallic half-quantized valley-like Hall effect, and the system in (f) realizes a quantized valley-like Hall effect, respectively. Here, the gapless Dirac cones are colored grey. The red and blue Dirac cones indicate that they gain local topological masses with the opposite signs. In (b)-(d), the blue and green surfaces represent ferromagnetic layers with magnetic moments oriented in opposite directions. In (e)-(f), the red surface corresponds to altermagnetic layers.}
\label{fig_illustration}
\end{figure}

In this work, we investigate the effect of ferromagnetism, antiferromagnetism, and altermagnetism on a WTI film. Our research uncovers an array of intriguing Hall effects facilitated by the interplay between magnetism and WTIs.  Moreover, in the magnetic WTI film, we find that the half-quantized Hall conductance is revealed through the integration of the Berry curvature over half of the Brillouin zone. This is different from the previous studies on the magnetic STI film, that the half-quantization is manifested through the integration of the Berry curvature over the whole Brillouin zone~\cite{Fu2022npjQM,ZouJY2022PRB,ZouJY2023PRB,Chen2023arXiv_SSQ,QinF2023PRB,QinF2022PRB}. Below, with close reference to Fig.~\ref{fig_illustration}, we provide a summary of these closely-related topological phases in the ferromagnetic and altermagnetic systems. It is noticed that the total Hall conductance is related to the number of magnetism-induced gapped surface Dirac cones.

(i) Figure~\ref{fig_illustration}(a): a WTI film without magnetization.  Each of the top and bottom surfaces hosts two gapless Dirac cones. Time-reversal symmetry ensures that the Hall conductance of this system is zero.

(ii) Figure~\ref{fig_illustration}(b): a metallic quantum anomalous Hall effect, a phenomenon that challenges conventional expectations and highlights the interplay of magnetism and the topology in WTIs.  The ferromagnetism is selectively introduced to the top surface of the WTI film.  The two Dirac cones on the top surface are gapped due to local time-reversal symmetry breaking caused by the magnetization. In contrast, the two Dirac cones on the bottom surface remain gapless, as this layer is not subjected to magnetization.
%

(iii) Figure~\ref{fig_illustration}(c): a quantum anomalous Hall effect with a higher Hall conductance plateau $\sigma_{xy}=2e^2/h$, resulting from contributions of $e^2/h$ from both top and bottom surfaces. Introducing ferromagnetism to both surfaces in the WTI film gaps all four Dirac cones. This scenario can be regarded as a double version of Case (ii), but it features two pairs of gapless chiral edge states within the insulating gap.

(iv) Figure~\ref{fig_illustration}(d): a quantized layer Hall effect with the top and bottom surfaces possessing quantized Hall conductances of opposite sign. Introducing ferromagnetism to both surfaces, with oppositely oriented magnetic moments, gaps all four Dirac cones. However, the two Dirac cones on the bottom surface acquire an opposite Dirac mass compared to those on the top.

(v) Figure~\ref{fig_illustration}(e): a metallic half-quantized valley-like Hall effect with a half-quantized Hall conductance $e^2/2h$ or $-e^2/2h$ distributed across different halves of the Brillouin zone. This effect is protected by the $C_4T$ symmetry, where $C_4$ is the 4-fold rotational symmetry and $T$ is time-reversal symmetry. Due to altermagnetism, the two Dirac cones at $X$ and $Y$ points on the top surface gain opposite masses, while those on the bottom surface remain gapless due to the lack of magnetization.

(vi) Figure~\ref{fig_illustration}(f): a quantized valley-like Hall effect with a quantized Hall conductance $e^2/2h$ or $-e^2/2h$ distributed across different halves of the Brillouin zone. This effect is protected by the $C_4T$ symmetry. The two Dirac cones at $X$ and $X'$ points, as well as those at $Y$ and $Y'$ points gain opposite masses due to the altermagnetism. This scenario can be regarded as a double version of Case (v),  but with two pairs of counter-propagating gapless chiral edge states within the insulating gap.

Additionally, we explore two distinct scenarios showcasing the antiferromagnetic effect depicted in Figs.~\ref{fig_illustration_AFM}(a) and \ref{fig_illustration_AFM}(b). We discover that these configurations lead to two distinct phenomena: a quantum anomalous Hall effect characterized by a higher Hall conductance plateau and a quantized layer Hall effect.

\section{Model and method}
\subsection{Model}
We start with a generic four-band topological insulator model defined on a cubic lattice~\cite{Zhang09np,LiuCX2012PE}
\begin{equation}
H=H_0(\bm{k})+H_\text{fm}(z)+H_\text{am}(k_x,k_y,z)+H_\text{afm}(z),
\label{Eq_Hamiltonian}
\end{equation}
\begin{equation}
H_0(\bm{k})=\mathcal{M}(\bm{k}) \Gamma_4+A (\Gamma_1 \sin k_x+\Gamma_2 \sin k_y+\Gamma_3 \sin k_z),
\end{equation}
where $\mathcal{M}(\bm{k})=M_0+6 B-2 B \sum_i \cos k_i, \Gamma_1=\sigma_1  \tau_1$, $\Gamma_2=\sigma_2  \tau_1, \Gamma_3=\sigma_3 \tau_1$,  and $\Gamma_4=$ $\sigma_0  \tau_3$. Here $M_0, B$ and $A$ are model dependent parameters. $\sigma$ and $\tau$ are Pauli matrices representing spin and orbital degrees of freedom, respectively.

In the absence of magnetization [i.e., $H_\text{fm}=0$, $H_\text{am}=0$, and $H_\text{afm}=0$], $H_0(\bm{k})$ represents a trivial insulator when $M_0>0$. When $M_0<-12 B$, it depicts an STI phase with the $Z_2$ index $(1 ; 000)$ or a STI phase with $(1 ; 111)$ when $0>M_0>-4 B$ or $-8 B>M_0>-12 B$.  It corresponds a WTI phase with $(0 ; 111)$ when $-4 B>M_0>-8 B$~\cite{FL07PRB}. This four-band model, originally proposed by Zhang~\emph{et al.}, has been successfully used to describe the STI of $\mathrm{Bi}_2 \mathrm{Se}_3$ family~\cite{Zhang09np}. As our primary focus is on the WTI phase, we concentrate on the parameter regime $-4 B>M_0>-8 B$.

The last three terms of the Hamiltonian in Eq.~\ref{Eq_Hamiltonian}, $H_\text{fm}(z)$, $H_\text{am}(k_x,k_y,z)$, and $H_\text{afm}(z)$ depict the layer-dependent effect of ferromagnetism, altermagnetism, and antiferromagnetism, respectively.
Specifically, the three types of magnetization have the following form
\begin{align}
H_\text{fm}(z)&=m(z)\sigma_3 \tau_0,\label{Eq_mag1}
\\
H_\text{am}(k_x,k_y,z)&=m'(z)(\cos k_x-\cos k_y)\sigma_3 \tau_0\label{Eq_mag2},
\\
H_\text{afm}(z)&=m''(-1)^z\sigma_3 \tau_0\label{Eq_mag3}.
\end{align}
Ferromagnetism can be induced through magnetic doping~\cite{Bernevig2022Nat,Tokura19nrp,Wang2021TI,Liu2021AdvMat}.  Equation~\eqref{Eq_mag2} represents a $d$-wave altermagnetic ordering ~\cite{Mejkal2020SA,Libor22PRX,Smejkal22PRX,Feng2022Natelec,Betancourt23PRL}, which can be achieved via the proximity effect. Equation~\eqref{Eq_mag3} depicts the antiferromagnetic ordering, where neighboring layers possess opposite magnetizations. The amplitude of the magnetization within each layer is $m''$.

The ferromagnetic term $m\left( z \right)$  and altermagnetic term $m^\prime\left( z \right)$  are given by
\begin{equation}
m\left( z \right) =\begin{cases}
	m_t\\
	0\\
	m_b\\
\end{cases}\begin{array}{l}
	z=1,2,3,\\
	\mathrm{elsewhere},\\
	z=n_z-2,n_z-1,n_z,\\
\end{array}
\end{equation}
and
\begin{equation}
m^\prime\left( z \right) =\begin{cases}
	m_t^\prime\\
	0\\
	m_b^\prime\\
\end{cases}\begin{array}{l}
	z=1,2,3,\\
	\mathrm{elsewhere},\\
	z=n_z-2,n_z-1,n_z.\\
\end{array}
\end{equation}
This is  more clearly illustrated in Fig.~\ref{fig_illustration}, where ferromagnetic and altermagnetic effects are confined solely to the top and bottom surface layers.

In the following content, we first introduce the nonmagnetic system in Sec.~\ref{Sec_Non}, then focus on the ferromagnetic and altermagnetic effects in Sec.~\ref{Sec_Fe}, and finally discuss the antiferromagnetic effect in Sec.~\ref{Sec_AFM}.

\subsection{Method}

The ferromagnetic and antiferromagnetic systems are characterized by the total Hall conductance
 \begin{equation}
\sigma_{xy}=\sum_{k_x,k_y,z}\sigma_{xy}(k_x,k_y,z),
\label{Eq_total_Hall}
\end{equation}
and the layer Hall conductance
\begin{equation}
\sigma_{xy}(z)=\sum_{k_x,k_y}\sigma_{xy}(k_x,k_y,z).
\label{Eq_layer_Hall_0}
\end{equation}
In Appendix~\ref{App_Hall}, we provide more details for calculating the layer- and momentum-resolved Hall conductance $\sigma_{xy}(k_x,k_y,z)$~\cite{Varnava18prb}. Here, the layer Hall conductance $\sigma_{xy}(z)$ counts the contributions to the total Hall conductance from the $z$-th layer. The total Hall conductance $\sigma_{xy}$ counts the contribution to the Hall conductance from all layers, i.e., $\sigma_{xy}=\sum_z\sigma_{xy}(z)$. Moreover, we define
\begin{align}
\sigma_{xy}^{t}=\sum_{z=n_z-3}^{z=n_z}\sigma_{xy}(z),
\sigma_{xy}^{b}=\sum_{z=1}^{z=4}\sigma_{xy}(z),
\label{Eq_layerHall}
\end{align}
which count the contributions to the Hall conductance from the top and bottom surface layers, respectively.

On the other hand, we adopt a momentum-resolved Hall conductance to characterize the altermagnetic system, with
\begin{equation}
\sigma_{xy}^{\prime}(\theta)=\sum_{k_x,k_y\in P(\theta),z}\sigma_{xy}(k_x,k_y,z),
\label{Eq_Hall_sigma}
\end{equation}
where $P(\theta)$ is the area enclosed by the green lines with a $\theta$ angle in Fig.~\ref{fig_AM}(a).
\section{Nonmagnetic system}
\label{Sec_Non}
\begin{figure}[t]
\centering
\includegraphics[width=\columnwidth]{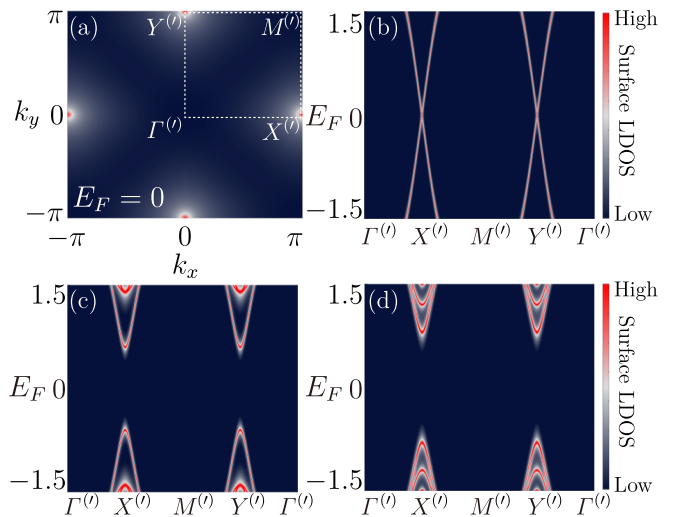}
\caption{(a) LDOS of the nonmagnetic surface as a function of $k_x$ and $k_y$ with the Fermi energy $E_F=0$. LDOS of the (b) nonmagnetic surface, (c) ferromagnetic surface, and (d) altermagnetic surface along the path of high symmetry points in the surface Brioullion zone [the white dashed line in (a)].}
\label{fig_spectrum}
\end{figure}

As depicted schematically in Fig.~\ref{fig_illustration}(a), the nonmagnetic system corresponds to a WTI film with four gapless Dirac cones. Two of these Dirac cones are situated at $X$ and $Y$ points of the top surface Brillouin zone, while the other two are located at $X^\prime$ and $Y^{\prime}$ points of the bottom surface Brillouin zone. This is further clarified in Figs.~\ref{fig_spectrum}(a) and \ref{fig_spectrum}(b), which show the local density of state (LDOS) of the nonmagnetic surface. The Hall conductance of the system is zero due to the existence of time-reversal symmetry $T=I_{n_z}\sigma_y \tau_0 K$, where $I_{n_z}$ is the identity matrix and $K$ denotes the complex conjugation.
\section{Ferromagnetic and altermagnetic systems}
\label{Sec_Fe}

In this section, we consider five different scenarios for the ferromagnetic system~[Figs.~\ref{fig_illustration}(b) and \ref{fig_illustration}(c)] and altermagnetic systems~[Figs.~\ref{fig_illustration}(e) and \ref{fig_illustration}(f)]. In the numerical calculations, the coefficients are given by the following tabular,
\begin{equation}%
\begin{tabular}
[c]{c|ccccccc}
 &$(\mathrm{i)}$&$(\mathrm{ii)}$&$(\mathrm{iii)}$
 &$(\mathrm{iv)}$& $(\mathrm{v)}$&$(\mathrm{vi)}$ \\\hline
$m_t$&	          0& V& V& V& 0 &0\\
$m_b$&	          0& 0& V& -V& 0 &0\\
$m_{t}^{\prime}$& 0& 0& 0& 0& V &V\\
$m_{b}^{\prime}$& 0& 0& 0& 0& 0 &V\\
\end{tabular}
\end{equation}%
The other parameters are taken as $A=1$, $B=1$, $M=-6$, and $V=0.6$.

\begin{figure}[t]
\centering
\includegraphics[width=\columnwidth]{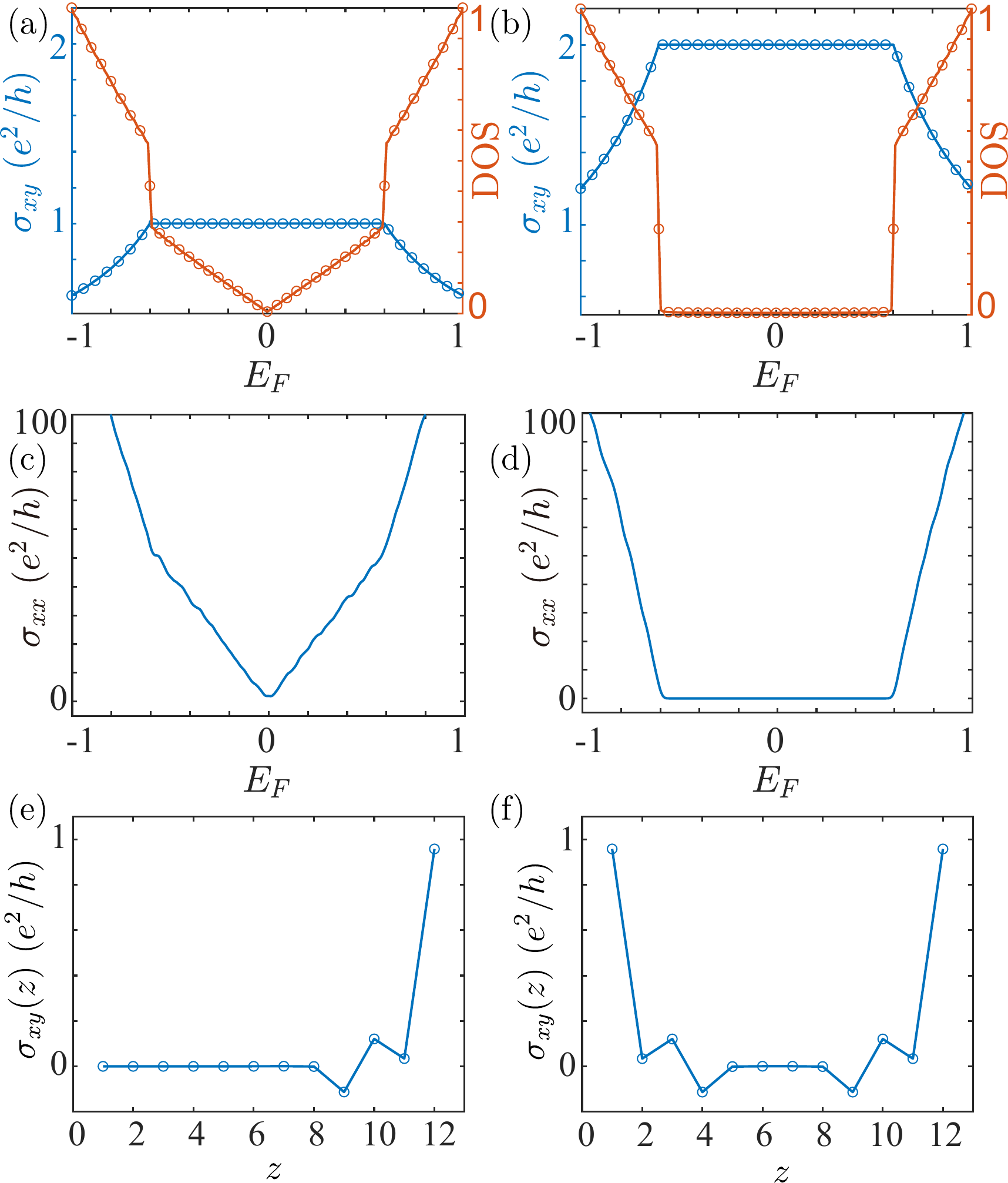}
\caption{(a) Numerically calculated Hall conductance $\sigma_{xy}$ (blue) and the DOS (red) as functions of the Fermi energy $E_F$ for the metallic quantum anomalous Hall effect shown in Fig.~\ref{fig_illustration}(b). (c) Numerically calculated longitudinal conductance $\sigma_{xx}$ as a function of the Fermi energy $E_F$ for the metallic quantum anomalous Hall effect. (e) The layer Hall conductance $\sigma_{xy}(z)$ of the metallic quantum anomalous Hall insulator as a function of the layer index $z$. (b), (d), and (f) are the same as (a), (c), and (e), except that they describe the quantum anomalous Hall effect with a higher plateau $\sigma_{xy}=2e^2/h$.}
\label{fig_semimag}
\end{figure}

\subsection{Metallic quantum anomalous Hall
effect without chiral edge states}
\label{Sec_MQAH}
Let us first consider the simplest case where the ferromagnetization is introduced only to the top surface layers of the WTI film [Fig.~\ref{fig_illustration}(b)].  The Dirac cones on the top surface are gapped out due to the breaking of local time-reversal symmetry induced by the magnetization~[Figs.~\ref{fig_illustration}(b) and~\ref{fig_spectrum}(c)]. The gapless Dirac cones on the bottom surface remain due to the absence of local magnetization~[Figs.~\ref{fig_illustration}(b) and~\ref{fig_spectrum}(b)].

Figure~\ref{fig_semimag}(a) shows the numerically calculated Hall conductance $\sigma_{xy}$ and the DOS as functions of the Fermi energy $E_F$. The system exhibits a quantized anomalous Hall conductance (the blue circle line) when the Fermi energy is inside the magnetic gap of the top surface ($\left|E_F\right|<V$). However, it is important to note that the system differs from conventional quantum anomalous Hall insulators, where the quantized Hall conductance only emerges when the Fermi energy resides within the insulating gap. In contrast, the present system lacks an insulating energy gap, as revealed by the numerical results of the DOS (the red circle line). Moreover, the DOS exhibits a linear increase as the Fermi energy increases, indicating its origin from the 2D gapless Dirac cones on the bottom surface.

Figure~\ref{fig_semimag}(c) shows the longitudinal conductance $\sigma_{xx}$ as a function of the Fermi energy $E_F$,  numerically obtained using the Kubo-Greenwood formula expressed in terms of the Chebyshev polynomials~\cite{Weibe06RMP,Carcia15PRL} (see Appendix~\ref{App_Long}). The system manifests a metallic quantum anomalous Hall effect, distinguished by a quantized Hall conductance and a nonvanishing longitudinal conductance. Figure~\ref{fig_semimag}(e) displays the layer Hall conductance as a function of layer index $z$ with $E_F=0$.  We find that $\sigma_{xy}^t=0.9998e^2/h$ and $\sigma_{xy}^b=0.0001e^2/h$ [see Eq.~\eqref{Eq_layerHall}]. The slight deviation from perfect quantization arises from the quantum confinement effect. These results further confirm that the quantized Hall conductance originates from the ferromagnetic top surface.

Very recently, the metallic quantum anomalous Hall effect has been proposed in an STI film with a ferromagnetic layer at interior layers~\cite{Bai2023arXiv}. Our work presents an alternative approach to realize the metallic quantum anomalous Hall effect. Moreover, the metallic quantum anomalous Hall effect can be regarded as a double version of the semi-magnetic topological insulator, which has been experimentally detected using the traditional six-terminal Hall-bar measurement and is characterized by a half-quantized Hall conductance and a nonvanishing longitudinal conductance~\cite{mogi2021experimental}. Thus, we believe that the metallic quantum anomalous Hall effect in our work could be observed using a similar experimental setup.

\subsection{Quantum anomalous Hall
effect with a higher plateau}
\label{Sec_QAH}
Here we consider the case shown in~Fig.~\ref{fig_illustration}(c), with parallel magnetization alignment on the top and bottom surfaces.  All the surface Dirac cones are gapped out due to the magnetic effect, and the top and bottom surfaces share the same LDOS shown in Fig.~\ref{fig_spectrum}(c). The system hosts a global energy gap characterized by a vanishing DOS when $\left|E_F\right|<V$ ~[see the red circle line in Fig.~\ref{fig_semimag}(b)]. This energy gap features a higher quantized Hall conductance $\sigma_{xy}=2e^2/h$~[the blue circle line in Fig.~\ref{fig_semimag}(b)]. Consequently, the system is identified as a quantum anomalous Hall insulator. The insulating nature of the system is further confirmed by calculating the longitudinal conductance $\sigma_{xx}$~[Fig.~\ref{fig_semimag}(d)], which remains zero as long as the chemical potential resides within the energy gap. Analysis of the layer Hall conductance reveals that both the top and bottom surfaces contribute a nearly quantized Hall conductance~[Fig.~\ref{fig_semimag}(f)], with $\sigma_{xy}^t=\sigma_{xy}^b=0.9998e^2/h$. In experiments, the quantum anomalous Hall effect can be detected using the traditional six-terminal Hall-bar measurement.

\subsection{Quantized layer Hall effect}
\label{Sec_QLHE}
\begin{figure}[t]
\centering
\includegraphics[width=\columnwidth]{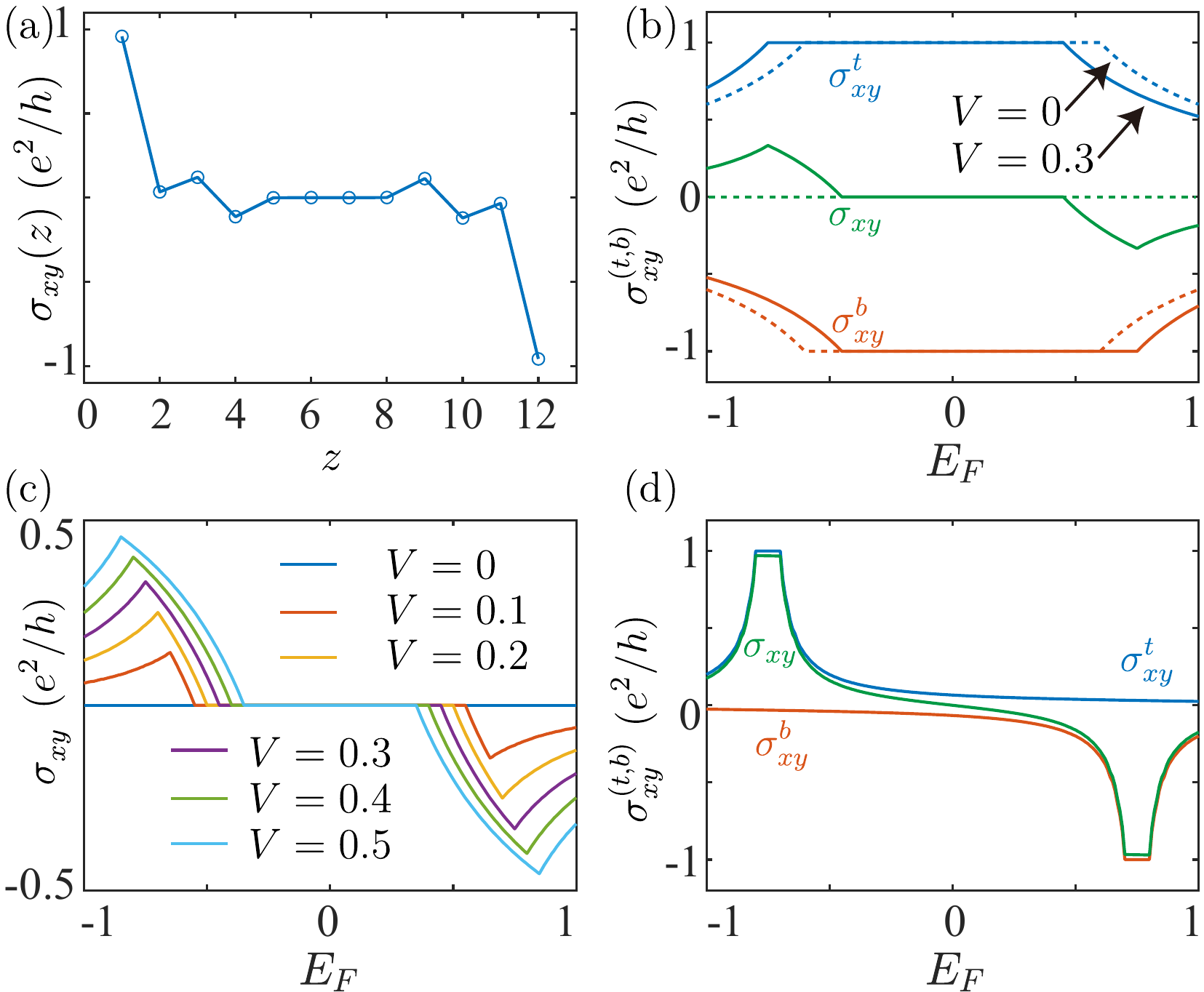}
\caption{(a) The layer Hall conductance $\sigma_{xy}(z)$ of the quantized layer Hall effect as a function of the layer index $z$. (b) $\sigma_{xy}$ (green), $\sigma_{xy}^t$ (blue), and $\sigma_{xy}^b$ (red) as functions of the Fermi energy. The dashed lines and solid lines correspond to $V=0$ and $V=0.3$, where $V$ is the strength of the external perpendicular electric field, respectively. (c) $\sigma_{xy}$ as a function of the Fermi energy for different $V$. (d) The same with (b), but with different film thickness $n_z=6$, electric field strength $V=1.5$, and the thickness of the magnetized layer is $1$. }
\label{fig_axion}
\end{figure}

\begin{figure}[th]
\centering
\includegraphics[width=1\columnwidth]{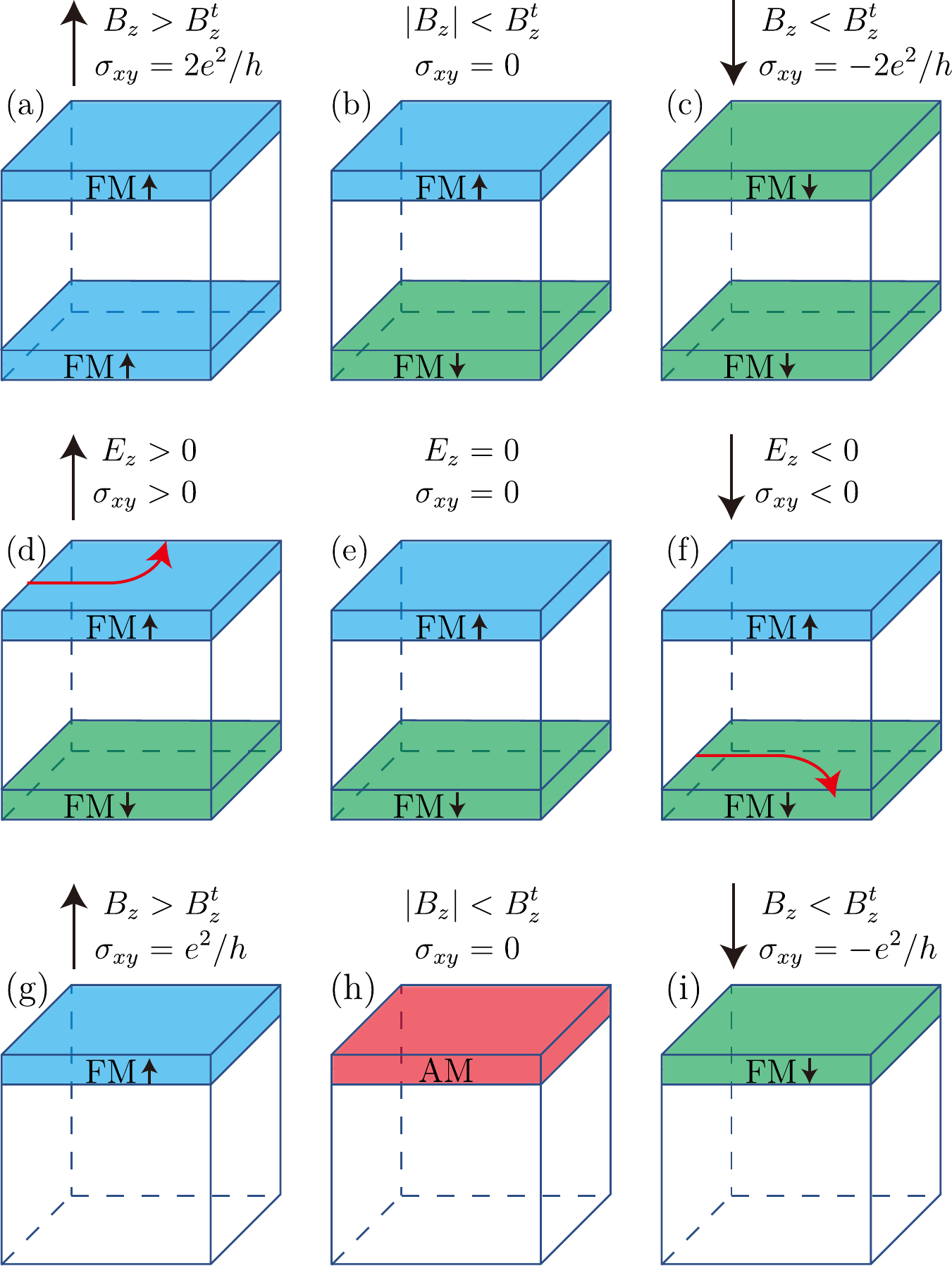}
\caption{[(a)-(f)] The quantized layer Hall effect~[see Sec.~\ref{Sec_QLHE}] can be detected by either applying a perpendicular magnetic field or a perpendicular electric field. By tuning the strength of the external magnetic field [see (a)-(c)], there would appear an intermediate zero Hall conductance plateau between the two quantized Hall conductance plateaus with $\sigma_{xy}=\pm 2e^2/h$. By tuning the strength of the external electric field [see (d)-(f)], there would appear a nonvanishing net Hall current on the surface layer [the red arrow line], with  its propagating direction depending on the direction of the electric field. (g)-(i) The metallic half-quantized valley-like Hall effect~[see Sec.~\ref{Sec_MHVHE}] can be detected by applying a perpendicular magnetic field. By tuning the strength of the external magnetic field [see (g)-(i)], there would appear an intermediate zero Hall conductance plateau between the two quantized Hall conductance plateaus with $\sigma_{xy}=\pm e^2/h$ and a nonvanishing longitudinal conductance $\sigma_{xx}$. Note that (a) and (c) correspond to the quantum anomalous Hall effect in Sec.~\ref{Sec_QAH}, and (g) and (i) correspond to the metallic quantum anomalous Hall effect described in Sec.~\ref{Sec_MQAH}. }
\label{fig_setup}
\end{figure}

Now we consider the case shown in~Fig.~\ref{fig_illustration}(d), where the magnetizations on the top and bottom surfaces are antiparallel. All the surface Dirac cones are gapped due to the magnetic effect, and the top and bottom surfaces share the same LDOS as shown in Fig.~\ref{fig_spectrum}(c). It is noticed that the top and bottom surfaces gain an opposite Dirac mass due to the antiparallel magnetization alignments on the top and bottom surfaces~[labeled by the red and blue gapped Dirac cones in Fig.~\ref{fig_illustration}(d)].

The breaking of local time-reversal symmetry results in a nonzero layer Hall conductance as shown in Fig.~\ref{fig_axion}(a). Moreover, the layers connected by the
$PT$ symmetry are exactly compensated, leading to a zero net Hall conductance. Here, $P=M\sigma_0\tau_z$ depicts the inversion symmetry, and $M$ is the orthogonal matrix that permutes the layers of the entire system perpendicularly. Figure~\ref{fig_axion}(b) shows $\sigma_{xy}^{t/b}$ and the total Hall conductance $\sigma_{xy}$ as functions of the chemical potential $E_F$ [see the dashed lines]. Both the top and bottom surfaces exhibit a quantized surface Hall conductance but with an opposite sign, establishing the quantized layer Hall effect.

The quantized layer Hall effect can be viewed as a double version of the axion insulator, which has been proposed for experimental detection by various approaches~\cite{Mogi17nm,Liu20nm,Gao2021Nature,ChenR2022NSR,Lei2023arXiv,ChenR21PRB,LiYH2022PRR}. One of the most direct approaches involves using the traditional transport measurement by applying an external magnetic field~[Figs.~\ref{fig_setup}(a)-\ref{fig_setup}(c)]~\cite{Mogi17nm,Liu20nm}. By varying the strength and direction of the external magnetic field, the Hall conductance of the system should fluctuate between $-2e^2/h$, $0$, and $2e^2/h$. The quantized layer Hall effect is characterized by the intermediate zero plateau, while the quantum anomalous Hall effect mentioned in Sec.~\ref{Sec_QAH} is characterized by the quantized Hall conductance $\pm 2e^2/h$.

Another method to detect the quantized layer Hall effect is by applying an external perpendicular electric field~[Figs.~\ref{fig_setup}(d)-\ref{fig_setup}(f)]~\cite{Gao2021Nature,ChenR2022NSR}, which breaks the $PT$ symmetry and induces an energy offset between the plateaus of the surface Hall conductance~[Fig.~\ref{fig_axion}(b)]. The electric-field-induced Hall conductance has opposite signs for opposite Fermi energies and its amplitude increases with the increasing strength of the field~[Fig.~\ref{fig_axion}(c)].  For moderate film thickness and strength of the external electric field, we find that the emergent Hall conductance can approximate the quantized value of $e^2/h$~[Fig.~\ref{fig_axion}(d)]. These signatures provide evidence for observing the quantized layer Hall effect in the ferromagnetic WTI films.

\begin{figure}[t]
\centering
\includegraphics[width=\columnwidth]{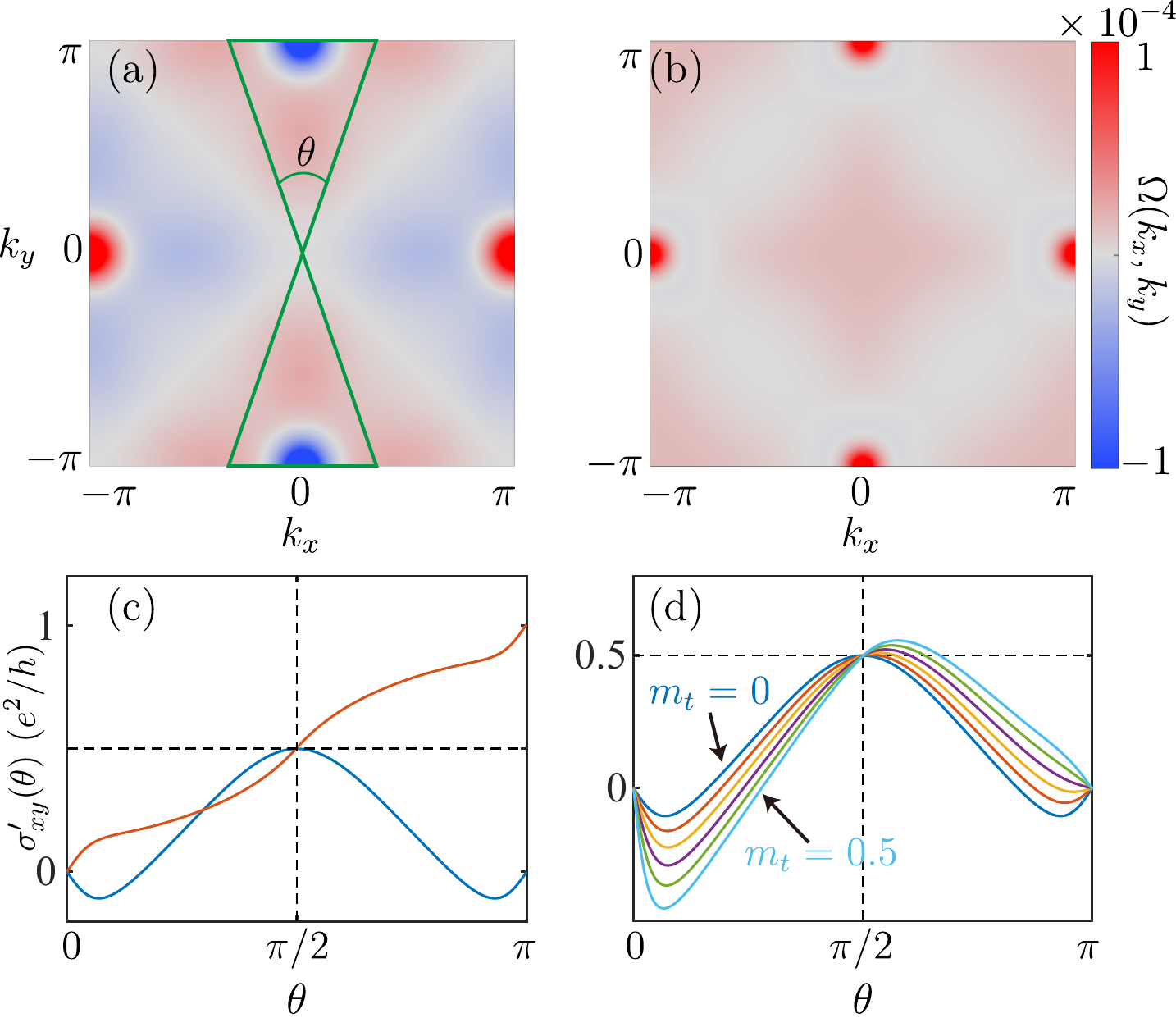}
\caption{(a) and (b) correspond to the Berry curvature distribution for the altermagnetic system shown in Fig.~\ref{fig_illustration}(e) and the ferromagnetic system shown in Fig.~\ref{fig_illustration}(b), respectively. (c) and (d) The momentum resolved Hall conductance $\sigma_{xy}^{\prime}(\theta)$ [see Eq.~\eqref{Eq_Hall_sigma}] as a function of $\theta$.  Here, $\theta$ corresponds to the angle between the two green lines in (a). In (c), the blue curve and red curve depict the altermagnetic case and the ferromagnetic case shown in (a) and (b), respectively. In (d), the curves depict the cases with a fixed altermagnetic strength and a varying ferromagnetic strength. More specifically, the parameters are given by (a) $m_t=0$ and $m_t^\prime=V$, (b) $m_t=V$ and $m_t^\prime=0$, (c) blue for $m_t=0$ and $m_t^\prime=V$ and red for $m_t=V$ and $m_t^\prime=0$, and (d) $m_t^\prime=V$ with $m_t$ ranging from $m_t=0$ to $m_t=0.5$.
}
\label{fig_AM}
\end{figure}

\subsection{Metallic half-quantized valley-like Hall effect}
\label{Sec_MHVHE}
We consider the altermagnetic case, which can be achieved by approximating the WTI to an altermagnetic material. The system breaks time-reversal symmetry but preserves the combined $C_4 T$ symmetry, where $C_4=I_{n_z} \exp(-i\sigma_z\tau_0\pi/4)R_4$ and $R_4$ is the rotational matrix with $R_4 (k_x,k_y)=(k_y,-k_x)$. The $C_4 T$ symmetry requires that \begin{equation}
\Omega(\bm{k})=-\Omega(R_4\bm{k}),
\label{Eq_Berry}
\end{equation}
which guarantees a zero total Hall conductance.

Figure~\ref{fig_AM}(a) shows the Berry curvature of the altermagnetic system as a function of $k_x$ and $k_y$. Our numerical results align with  Eq.~\eqref{Eq_Berry}, where the Berry curvature $\Omega(\bm{k})$ is opposite in sign to $\Omega(R_4\bm{k})$.
The blue curve in Fig.~\ref{fig_AM}(c) shows the Hall conductance $\sigma_{xy}^{\prime}(\theta)$ as a function of $\theta$, where $\sigma_{xy}^{\prime}(\theta)$ is the total Hall conductance contributed by the $k$ points within the green cone area shown in Fig.~\ref{fig_AM}(a). We observe that half of the Brillouin zone contributes to a half-quantized Hall conductance. Consequently, the other half of the Brillouin zone must contribute another half-quantized Hall conductance with the opposite sign due to the $C_4 T$ symmetry. Moreover, similar to the situation discussed in Sec.~\ref{Sec_MQAH}, the bottom surface is gapless, making the system a metal. Therefore, we identify this system as a metallic half-quantized valley-like Hall effect.

For a comparative study, we also examine the Berry curvature distribution of the ferromagnetic case [see Sec.~\ref{Sec_MQAH} and Fig~\ref{fig_illustration}(b)]. The ferromagnetic system is protected by the $C_4$ symmetry and requires that $\Omega(\bm{k})=\Omega(R_4\bm{k})$. Thus, each half of the Brillouin zone will contribute a half-quantized Hall conductance, resulting in a quantized Hall conductance overall. This implies that each gapped Dirac cone is associated with a half-quantized Hall conductance, and the half-quantization is manifested through half of the Brillouin zone in the magnetic WTI film. This scenario differs from previous studies on magnetic STI films, where the half-quantization in the semimagnetic topological insulator phase is manifested through the integration of the Berry curvature over the entire Brillouin zone~\cite{Fu2022npjQM,ZouJY2022PRB,ZouJY2023PRB,Chen2023arXiv_SSQ}.

Furthermore, Figure~\ref{fig_AM}(d) shows $\sigma_{xy}^{\prime}(\theta)$ as a function of $\theta$, in a system where both the ferromagnetic and altermagnetic effects coexist. We find that the relationship $\sigma_{xy}(\pi/2)= 0.5$ is stable by varying the strength of ferromagnetic ordering.

\subsection{Quantized valley-like Hall effect}
\begin{figure*}[ht]
\centering
\includegraphics[width=1.8\columnwidth]{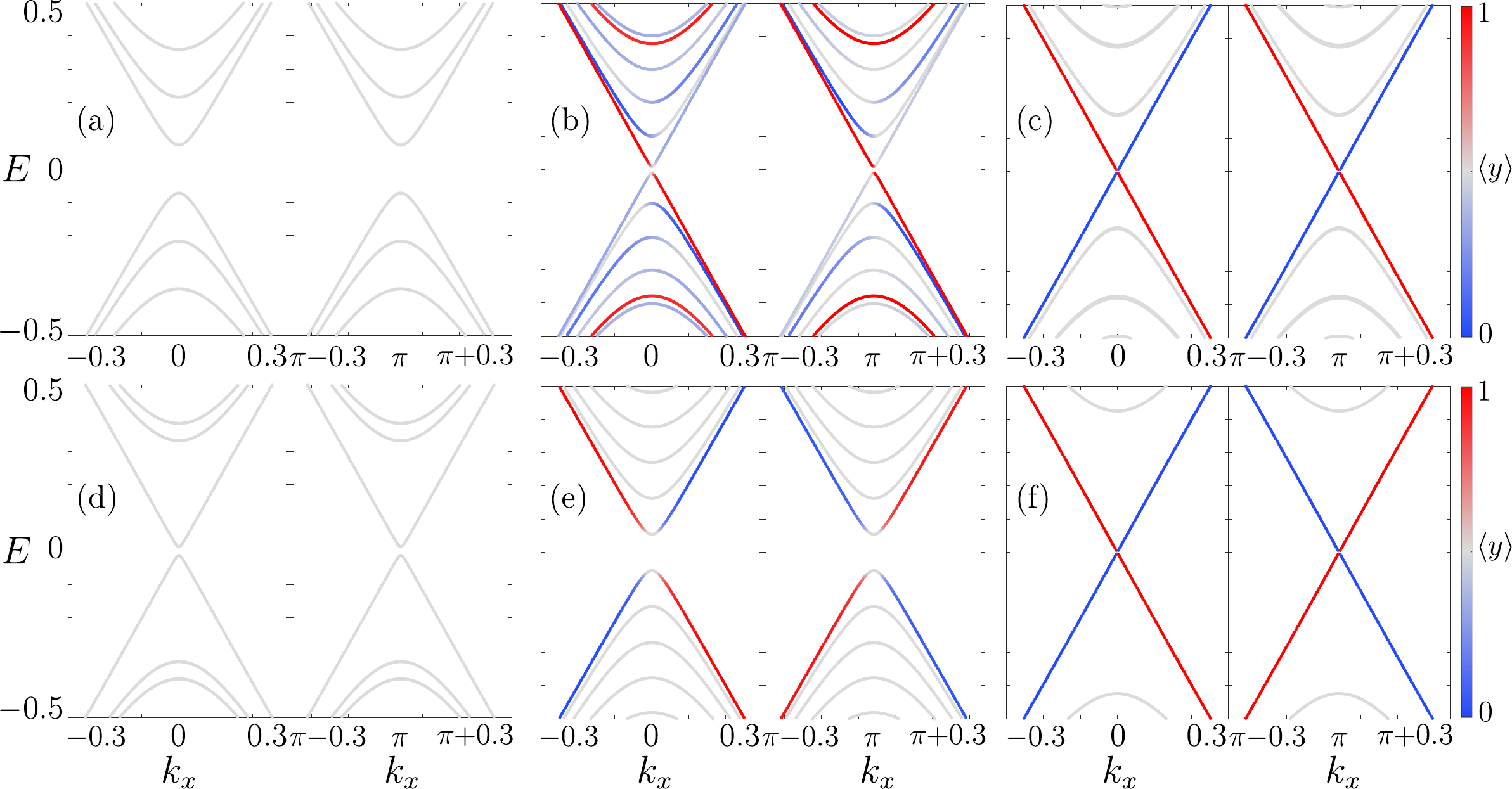}
\caption{Energy spectrum of the six systems, with periodic boundary condition along the $x$-direction and open boundary conditions along the $y$- and $z$-directions. The color scheme indicates the wave function distribution along the $y$-direction. Here, for ease of visualization, only the spectra near $k_x=0$ and $k_x=\pi$ are shown. The system size is taken as $n_y=20$ and $n_z=12$.}
\label{fig_spectrum_open}
\end{figure*}

Now we consider the case shown in Fig.~\ref{fig_illustration}(f), where the altermagnetic effect is introduced to both the top and bottom surface layers. The situation in Fig.~\ref{fig_illustration}(f) closely resembles that in Fig.~\ref{fig_illustration}(e). However, there are three key distinctions: (1) The Berry curvature and the corresponding Hall conductance double. In other words, half of the Brillouin zone contributes a quantized Hall conductance, and the other half contributes another quantized Hall conductance with the opposite sign; (2) the system is an insulator because both the top and bottom surfaces are gapped; and (3) the system hosts a chiral edge state inside the insulating surface gap [see Fig.~\ref{fig_spectrum_open} and Appendix~\ref{App_Spectrum}]. Consequently, the system is identified as a quantized valley-like Hall effect.

\section{Antiferromagnetic system}
\label{Sec_AFM}
\begin{figure}[tpb]
\centering
\includegraphics[width=\columnwidth]{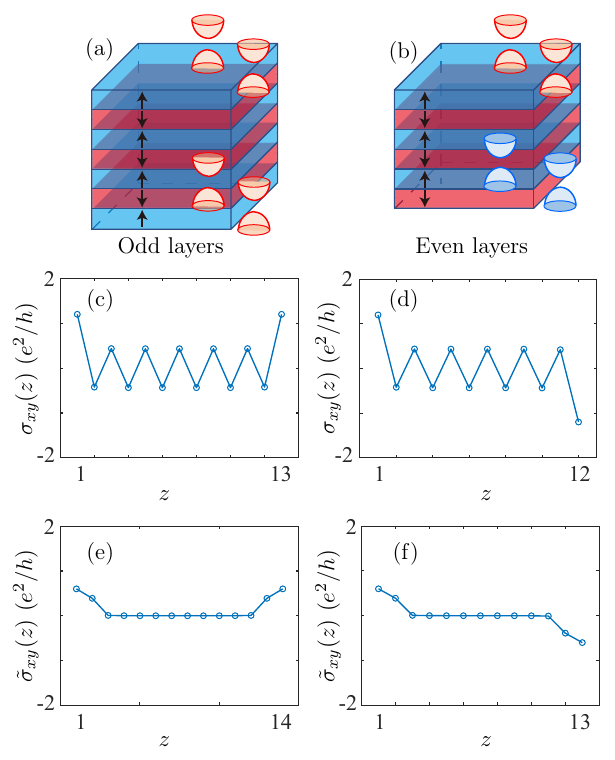}
\caption{(a)-(b) Schematic illustrations of two different antiferromagnetic configurations with odd and even layers, respectively. In both systems, all the four Dirac cones are gapped out due to the antiferromagnetization. Here, the red and blue Dirac cones indicate that they gain local topological masses with the opposite sign. The red and blue layers represent ferromagnetic layers with magnetic moments oriented in opposite directions. (c)-(d) The layer Hall conductance $\sigma_{xy}(z)$ and (e)-(f) the renormalized layer Hall conductance $\tilde{\sigma}_{xy}(z)$ as functions of the layer index $z$. In the numerical calculations, we take $m''=0.6$.}
\label{fig_illustration_AFM}
\end{figure}

In this section, we investigate the antiferromagnetic effect on WTI films,  focusing on the two scenarios depicted in Figs.~\ref{fig_illustration_AFM}(a) and \ref{fig_illustration_AFM}(b). In the presence of antiferromagnetism, the Dirac cones on the top and bottom surfaces acquire local topological masses with  the same sign for odd layers [Fig.~\ref{fig_illustration_AFM}(a)] and the opposite signs for even layers [Fig.~\ref{fig_illustration_AFM}(b)]. Figures~\ref{fig_illustration_AFM}(c) and \ref{fig_illustration_AFM}(d)  illustrate the layer Hall conductance $\sigma_{xy}(z)$  as a function of the layer index $z$. Each magnetic layer exhibits a nonzero layer Hall conductance with its sign depending on the direction of its magnetic moment. Additionally, the layer Hall conductance is enhanced at the surface layers due to the surface Dirac cones. In the odd-layer system, the total Hall conductance is $2e^2/h$ due to the breaking of time-reversal symmetry. In the even-layer system, the total Hall conductance is always zero protected by the $PT$ symmetry.

To mitigate the oscillatory behavior of the bulk layers, we define the renormalized Hall conductance~\cite{Varnava18prb} as
\begin{equation}
\tilde{\sigma}_{xy}(z)=\frac{\sigma_{xy}(z)+\sigma_{xy}(z+1)}{2},
\end{equation}
where $z=2,\cdots,n_z$, $\tilde{\sigma}_{xy}(1)=\sigma_{xy}(1)/2$, and $\tilde{\sigma}_{xy}(n_z+1)=\sigma_{xy}(n_z)/2$. This approach effectively counts each layer once, except for the surface layers, which are counted with a weight of $1/2$. It can be considered an application of the sliding window averaging method~\cite{Varnava18prb}.

Figures~\ref{fig_illustration_AFM}(e) and \ref{fig_illustration_AFM}(f) show the renormalized layer Hall conductance $\tilde{\sigma}_{xy}(z)$ as a function of layer index $z$. The Hall conductance contributed by the bulk layers is compensated to zero due to the local $PT$ symmetry at the bulk. Moreover, we obtain $\sum_{z=1}^{z=3}\tilde{\sigma}_{xy}(z)=0.9980 e^2/h$ for the top surface layers of both systems, and $\sum_{z=n_z-2}^{z=n_z}\tilde{\sigma}_{xy}(z)=(-) 0.9980 e^2/h$ for the odd-layer (even-layer) systems. By adopting the renormalized layer Hall conductance, we extract the quantized surface Hall conductance from the bulk layers.  This indicates that the system in Fig.~\ref{fig_illustration_AFM}(a) realizes a quantum anomalous Hall effect with a higher Hall conductance plateau, while the system in Fig.~\ref{fig_illustration_AFM}(b) realizes a quantized layer Hall effect.

\section{Conclusion}
In this study, we investigate the magnetic effects on films of WTIs. Our findings demonstrate that the interplay between magnetism and WTIs gives rise to a diverse range of anomalous Hall effects, including the metallic quantum anomalous Hall effect without chiral edge states, the quantum anomalous Hall effect with a higher Hall conductance plateau, the quantized layer Hall effect, the metallic half-quantized valley-like Hall effect, and the quantized valley-like Hall effect. Notably, the quantized layer Hall effect, metallic half-quantized valley-like Hall effect, and quantized valley-like Hall effect are not observed in magnetic STIs.

\begin{acknowledgments}
D.-H.X. was supported by the NSFC (under Grant Nos.~12074108, 12474151 and 12347101), the Natural Science Foundation of Chongqing (Grant No.~CSTB2022NSCQ-MSX0568). R.C. acknowledges the support of the NSFC (under Grant No. 12304195) and the Chutian Scholars Program in Hubei Province.  B.Z. was supported by the NSFC (under Grant No. 12074107), the program of outstanding young and middle-aged scientific and technological innovation team of colleges and universities in Hubei Province (under Grant No. T2020001) and the innovation group project of the natural science foundation of Hubei Province of China (under Grant No. 2022CFA012).
\end{acknowledgments}

{\color{blue}\emph{Note-added}.}---
Recently, we became aware of a complementary study, which focus on a two-band two-dimensional topological semimetal, where the Brillouin zone is divided into two patches characterized by half-quantized Berry curvature fluxes with opposite signs due to the presence of the $C_{4z} T$ magnetic symmetry~\cite{ZouJy2024CP}.

\appendix

\section{Hall conductance}
\label{App_Hall}
In the calculations, the Hall conductances in Eqs.~\eqref{Eq_total_Hall}-\eqref{Eq_Hall_sigma} are calculated from the following expression~\cite{Varnava18prb}:
\begin{equation}
\sigma_{xy}(k_x,k_y,z)=\frac{-4 \pi e^2}{N_kSh} \operatorname{Im}  \sum_{v v^{\prime} c } X_{v c } Y_{v^{\prime} c }^{\dagger} \rho_{v v^{\prime} }(z),
\end{equation}
which corresponds to the contribution to the total Hall conductance $\sigma_{xy}$ from the $z$-th layer of a certain $(k_x,k_y)$ point. The matrix element for the position operator along the $x$ or $y$ directions, is denoted as $X(Y)_{v c }=\left\langle\psi_{v }|x(y)| \psi_{c }\right\rangle=$ $\frac{\left\langle\psi_{v }\left|i \hbar v_x\left(v_y\right)\right| \psi_{c }\right\rangle}{E_{c }-E_{v }}$, which is related to the energy difference between the conduction and valence bands $E_{c }-E_{v }$. The indices $v$ and $c$ represent the valence and conduction bands. $\rho_{v v^{\prime} }(z)$ is the projection matrix on to the corresponding $z$-th layer, which implies a summation over all orbitals $v, v^{\prime}, c$ belonging to that layer. $N_k$ represents the number of $k$-points and $S$ represents the unit cell area.


\section{Longitudinal conductance}
\label{App_Long}
The longitudinal conductance is obtained by using the Kubo-Greenwood formula expressed in terms of the Chebyshev Polynomials~\cite{Weibe06RMP,Carcia15PRL}. In zero temperature, it can be expressed as~\cite{Carcia15PRL}
\begin{equation}
\sigma_{xx}\left(\tilde{E}_F\right)=\frac{4 e^2 \hbar}{\pi S} \sum_{m, n \leq m} \mu_{n m}^{xx} T_n\left(\tilde{E}_F\right) T_m\left(\tilde{E}_F\right),
\end{equation}
where
\begin{equation}
\mu_{m n}^{xx} \equiv \frac{g_m g_n}{\left(1+\delta_{n 0}\right)\left(1+\delta_{m 0}\right)} \operatorname{Tr}\left[v_x T_m(\tilde{H}) v_x T_n(\tilde{H})\right],
\end{equation}
and
\begin{equation}
g_m^J=\frac{(M-m+1) \cos \frac{\pi m}{M+1}+\sin \frac{\pi m}{M+1} \cot \frac{\pi}{M+1}}{M+1}
\end{equation}
is the Jackson Kernel, $T_m(x)=\cos(m\arccos x)$ is the Chebyshev polynomials of the first kind. $\tilde{H}$ is the rescaled Hamiltonian so that its eigenvalues $\tilde{E}_F$ is contained in the interval $\left[-1,1\right]$.

In the numerical calculation for the longitudinal conductance, the system size is taken as $400\times 400$, the moment $M=1000$, the disorder strength $W=0.1$ and the results are obtained after averaging on 20 independent disorder configurations.

\section{Spectrum with open boundary conditions along the $y$- and $z$-directions}
\label{App_Spectrum}
In addition, we plot the energy spectrum of the six systems, with periodic boundary condition along the $x$-direction and open boundary conditions along the $y$- and $z$-directions. For Case (i) shown in Fig.~\ref{fig_spectrum_open}(a), there is no chiral current due to the absence of magnetism. The systems in Cases (ii) and (iii) break the $T$ symmetry, which allow chiral current propagating along the same direction near $k_x=0$ and $k_x=\pi$ points. In Case (iv), each band is double degenerate and there is no net chiral current for the degenerate bands due to the $PT$ symmetry. In Cases (v) and (vi), there appear chiral currents propagating along the opposite directions $k_x=0$ and $k_x=\pi$ points, which results in a vanishing net chiral current.

Furthermore, it is noticed that the energy gaps between the edge states in Figs.~\ref{fig_spectrum_open}(b) and \ref{fig_spectrum_open}(e) are attributed to the quantum confinement effect. For a slab system with a large enough $n_y$, the system corresponds to a metal due to the gapless Dirac cone on the bottom surface layers. The situation is different from the case in Figs.~\ref{fig_spectrum_open}(c) and \ref{fig_spectrum_open}(f), with an insulating bulk even in the large $n_y$ limit.

\bibliographystyle{apsrev4-1-etal-title_6authors}
\bibliography{../refs-transport}


\end{document}